\documentclass[12pt,preprint]{aastex}

\usepackage{wasysym}
\newcommand\moon{{m}}

\newcommand\Mm{M_{\moon}}
\newcommand\Rea{{R_{\oplus}}}
\newcommand\Mea{{M_{\oplus}}}
\newcommand\K{\,{\rm K}}

\newcommand\gcm{{\rm\,g\,cm^{-3}}}
\newcommand\eV{{\rm\,eV}}
\newcommand{\G}{{\rm\, G}}
\newcommand\dyn{{\rm\,dyn}}

\newcommand\cm{{\rm\,cm}}
\newcommand\km{{\rm\,km}}
\newcommand\gm{{\rm\,g}}
\newcommand\<{{\langle}}
\renewcommand\>{{\rangle}}

\shortauthors{Gammie et al.}
\shorttitle{}

\title{A hot big bang theory: magnetic fields and the early evolution of the protolunar disk}

\author{C. F. Gammie}
\affil{Department of Astronomy, University of Illinois, 1002 West Green Street, Urbana, IL, 61801}

\author{Wei-Ting Liao}
\affil{Department of Astronomy, University of Illinois, 1002 West Green Street, Urbana, IL, 61801}

\author{P. M. Ricker}
\affil{Department of Astronomy, University of Illinois, 1002 West Green Street, Urbana, IL, 61801}

\begin{document}

\begin{abstract}

The leading theory for the formation of the Earth's moon invokes a collision between a Mars-sized body and the proto-Earth to produce a disk of orbiting material that later condenses to form the Moon.  Here we study the early evolution of the protolunar disk.  First, we show that the disk opacity is large and cooling is therefore inefficient ($t_{cool}\Omega \gg 1$).  In this regime angular momentum transport in the disk leads to steady heating unless $\alpha < (t_{cool}\Omega)^{-1} \ll 1$. Following earlier work by Charnoz and Michaut, and Carballido et al., we show that once the disk is completely vaporized it is well coupled to the magnetic field.  We consider a scenario in which turbulence driven by magnetic fields leads to a brief, hot phase where the  disk is geometrically thick, with strong turbulent mixing.  The disk cools by spreading until it decouples from the field.  We point out that approximately half the accretion energy is dissipated in the boundary layer where the disk meets the Earth's surface.  This creates high entropy material close to the Earth, driving convection and mixing.  Finally, a hot, magnetized disk could drive bipolar outflows that remove mass and angular momentum from the Earth-Moon system.

\end{abstract}

\section{Introduction}

The giant impact theory for the origin of the Earth's moon invokes a collision with a Mars-sized impactor  \citep{Hartmann1975, Cameron1976}.  Such collisions are expected to be common in young planetary systems \citep[e.g.][]{Chambers1998, Kenyon2006, Meng2014}.  The giant impact has been modeled numerically \citep[e.g.][]{Benz1985, Canup2004, Wada2006, Canup2013, Nakajima2014}.  It typically leads to the formation of a circumterrestrial disk and, in the giant impact scenario, the disk eventually condenses to form the Moon at a radius comparable to the Roche radius $\simeq 2.9 \Rea$.  

The initial conditions for the giant impact are characterized by a minimum of 10 parameters, including two masses, a relative velocity in the plane of the collision, and the spin angular momentum vector of each body. The two bodies may also differ in their chemical composition, isotopic composition, and magnetic field strength and geometry. While a great deal is now understood about the outcome of these collisions, the collision parameters remain uncertain.  

Early simulations were constructed under the assumption that the angular momentum of the system was approximately conserved from the impact to the present day \citep[e.g.][]{Canup2004}, when tidal coupling has transferred most of the angular momentum of the Earth-Moon system to the Moon's orbit.  Later work has challenged this assumption.  In particular, the Earth-Moon system may pass through an ``evection'' resonance where the Moon's apsidal precession period is close to one year.  The resulting resonant coupling then removes angular momentum from the Earth-Moon system \citep{Cuk2012}.

Uncertainties in the initial conditions for the impact translate into uncertainties in physical conditions in the post-impact protolunar disk. \cite{Nakajima2014}, for example, consider post-impact disks that vary widely in their mass, angular momentum, and entropy (and hence vapor fraction).  A common outcome, however, is a disk with mass $M_D \simeq 2 \Mm$, typical radius $(L_D/M_D)^2/(G \Mea) \simeq 2.5 \Rea$ ($L_D \equiv$ total disk angular momentum), and a distribution of temperatures from $3000-7000\K$.\footnote{Temperatures tend to be lower in models that assume a slow spinning Earth and angular momentum conservation \citep[e.g.][]{Canup2004} and higher in models with an initially fast-spinning Earth with later resonant angular momentum removal \citep{Cuk2012}; see \cite{Nakajima2014}.}

The chemical and isotopic composition of the present-day Earth and Moon potentially provide strong constraints on giant impact models \citep[e.g.][]{Jones2000, Wiechert2001, Dauphas2014, Melosh2014}.  Earth differs sharply in chemical composition from the Moon, and in particular has a substantial iron core.  The bulk lunar density $\rho_{\moon} \approx 3.34 \gcm$  \citep{Bills1977} compares to $\rho_\oplus \approx 5.5 \gcm$, which implies the Moon has a small iron core with $1-10\%$ of its mass, in contrast to Earth's iron core, which contains $\approx 30\%$ of its mass \citep{Canup2004b}.

The Earth and Moon are surprisingly similar in isotopic abundance, however, in light of the differences between the Earth and Mars and between Earth and some meteorites.  The lunar oxygen isotope ratio ($\delta ^{17} {\rm O} / \delta ^{18} {\rm O}$) lies very close to the terrestrial fractionation line, but far from Mars and other solar system bodies.  A similar trend is also found in other elements, including refractory elements such as Ti. 

There are at least two ways of producing a similar isotopic composition in the Moon and the Earth \citep[see review of][]{Melosh2014}: mix material between the Earth's mantle and the protolunar disk either after the impact \citep[e.g.][]{Pahlevan2007} or during the impact \citep{Cuk2012}, or invent a scenario in which the impactor and the Earth begin with nearly identical  isotopic composition \citep{Belbruno2005,Mastro2015}.

Our ability to assess the consequences of the giant impact and its aftermath relies on numerical models of the impact.  In this paper we ask whether treating the impact and post-impact disk using an ideal hydrodynamics model is self-consistent.  In \S 2 we introduce a reference disk model, evaluate its opacity, and show that cooling is inefficient, so that unless $\alpha$ is very small the disk will experience runaway heating.  In \S 3 we evaluate the conductivity of a hot vapor disk and show that it is well coupled to the magnetic field.  In \S 4 we investigate implications of magnetic coupling for development of magnetorotational instability driven turbulence.  In \S 5 we describe a scenario in which the disk and boundary layer are strongly magnetized and heat up to the virial temperature, producing rapid accretion, spreading, mixing, and potentially outflows.  \S 6 contains a summary and discussion.  

\section{Disk thermal evolution}

In the course of the giant impact material is shock heated and lofted into orbit around the Earth.  This material may be solid, liquid, or vapor.  Most simulations of the collision have at least 10\% vapor fraction \citep{Canup2004}; in some nearly the entire disk is vapor \citep{Wada2006,Nakajima2014}.  The temperature of the postimpact disk depends on the equation of state: the collision energy per nucleon at $11 \km \sec^{-1}$ is $\sim 0.6 \eV$, the dissociation energy per nucleon (for SiO) is $\sim 0.2 \eV$, and the latent heat of vaporization per nucleon (for Fe) is $\sim 0.06 \eV$, so dissociation and latent heat are not negligible.  \footnote{As another example, \cite{Fegley2012} evaluate the energy required to vaporize forsterite at 1 bar, beginning at $300\K$: $\approx 0.09 \eV$ per nucleon.} Since the coupling to magnetic field is exponentially sensitive to temperature (hotter plasma is better coupled) we begin by investigating the thermal evolution of the disk.

\subsection{Reference disk}

For definiteness, consider a reference disk with vapor mass $m \Mm$ and surface density 
\begin{equation}
\Sigma = \Sigma_0 \, e^{-r/r_0}
\end{equation}
where $r \equiv x \Rea, r_0 \equiv x_0 \Rea$, $\Sigma_0 \simeq m \Mm/(2 \pi r_0^2)$, and $\Mm \equiv$ mass of the Moon.  This choice is motivated by inspection of simulation results \citep[e.g.][]{Canup2013}.  The vapor disk may overlie a thin midplane disk containing liquids and solids at lower entropy; for now we will assume that the vapor is hot enough that it is not mixed with liquid, but return to consider a mixed liquid/vapor disk later.  Then 
\begin{equation}
\Sigma = 2.9 \times 10^7 \, \frac{m}{x_0^2} \, e^{-x/x_0}\, \gm \cm^{-2}
    \rightarrow 1.2 \times 10^6 \, \gm \cm^{-2}.
\end{equation}
Here and below, the expression following the arrow applies to a reference model with $m = 5$ at a fiducial location $x = x_0 = 3$ which we will also assume in numerical estimates.  About 70\% of the disk mass lies between $x_0 e^{-1} < x < x_0 e^1$, in conditions not far from this reference model.  Since disk temperature vary sharply, however, we retain the temperature dependence.

From hydrostatic equilibrium, the disk scale height
\begin{equation}
\frac{H}{r} = 0.15 \, T_5^{1/2} \, x^{1/2},
\end{equation}
where $T_5 \equiv T/5000\K$.  This assumes the disk is thin and therefore in Keplerian orbits, and the mean molecular weight $\mu = \mu_{SiO} \approx 44 m_p$.  Then density 
\begin{equation}
\rho \simeq \frac{\Sigma}{2 H} 
= 0.16 \, m T_5^{-1/2} x^{-3/2} x_0^{-2} e^{-x/x_0}\, \gm \cm^{-3} 
\, \rightarrow 6.1 \, \times 10^{-3} \, T_5^{-1/2} \,\gm \cm^{-3}
\end{equation}
This implies number density
\begin{equation}
n \simeq 2 \times 10^{21} \, m T_5^{-1/2} x^{-3/2} x_0^{-2} e^{-x/x_0}\, \cm^{-3}
\, \rightarrow \, 8.3 \, \times 10^{19} \, T_5^{-1/2} \,\cm^{-3}
\end{equation}
and pressure 
\begin{equation}
p = 1.5 \times 10^{9} \, m T_5^{1/2} x^{-3/2} x_0^{-2} e^{-x/x_0} \dyn \cm^{-2} \, \rightarrow \,
5.8 \times 10^7 \, T_5^{1/2} \, \dyn \cm^{-2},
\end{equation}
or $\simeq 58$ bar.

\subsection{Heating}

In standard thin disk theory the turbulent shear stress $w_{r\phi}$ is characterized by the dimensionless parameter $\alpha = w_{r\phi}/p$ \citep{Shakura1973}.  This is equivalent to adopting a turbulent kinematic viscosity $\nu \simeq \alpha c_s H$.  The heating rate per unit area
\begin{equation}
Q \simeq \alpha \Omega \int dz \, p
\end{equation}
The heating timescale is
\begin{equation}
t_{heat} = \frac{1}{Q}\int dz \, u
\end{equation}
where $u$ is the internal energy.  Then
\begin{equation}
t_{heat}\Omega \sim \alpha^{-1}
\end{equation}
for an ideal gas model.  

What is $\alpha$ for the protolunar disk?  If the disk is magnetically coupled then there is the possibility that the magnetorotational instability \citep[MRI, ][]{Balbus1991} drives magnetohydrodynamic turbulence \citep[e.g.][]{Hawley1995}.  Saturation of the MRI is not fully understood.  In numerical experiments the average $\alpha$ is known to depend on background field strength, viscosity, resistivity, and stirring of the disk by convection.  Nevertheless, simulations  of MRI-driven turbulence commonly measure $\alpha \sim 10^{-2}$ \citep[see, e.g.,][]{Turner2014}. 

Let us suppose, however, that the disk is initially too cool to couple to the magnetic field, and that the MRI is absent. Then gravitational instabilities \citep{Thompson1988, Ward2012, Ward2014, Gammie2001}, zombie vortex instabilities \citep{Marcus2015}, the subcritical baroclinic instability \citep[see][although the protolunar disk cooling time is likely too long]{Lesur2010, Klahr2003}, vertical shear instabilities \citep{Nelson2013,Richard2016}, turbulence associated with rain-out (liquid phase settling toward the midplane), and externally driven density waves, may contribute to $\alpha$.

Other sources of heat cannot be modeled as a turbulent viscosity.  The disk extends inwards to Earth's surface, where there is a shear (boundary) layer between the pressure-supported planet and the rotationally supported disk.  The supersonic shear layer is unstable to sound waves \citep{Belyaev2012a}, and these give rise to radially propagating spiral shocks \citep{Belyaev2012a, Belyaev2013a, Belyaev2013b}.  The energy per unit mass dissipated in the boundary layer is $(1/2) \Rea^2 (\Omega_K^2 - \Omega_\oplus^2)$ ($\Omega_K^2 \equiv G \Mea/\Rea^3$; $\Omega_\oplus \equiv$ Earth's spin frequency).  This is plausibly comparable to the total orbital kinetic energy, since if all the present-day angular momentum in the Earth-Moon system were placed in Earth's spin, the spin period would be $4 hrs$, which is much longer than low-Earth orbit period of $84$ min \citep[but see][]{Cuk2012}.

In a mixed liquid/vapor disk, settling of liquid drops can heat the vapor.  For a fixed liquid fraction $f$ the energy released when liquid in an initially well-mixed disk settles to the midplane is $(1/2) f \Sigma H^2 \Omega^2$, comparable to the total thermal energy for $f \sim 1/2$.  This energy is available on the settling timescale $\tau_{sett}$, which depends on drop size $a$.  

Raindrop radii grow until disruptive aerodynamic forces at terminal velocity, $\sim \rho_L a^3 g$ ($g \equiv$ gravitational acceleration; $\rho_L \equiv$ liquid density) are comparable to surface tension force $\sim \sigma a$ ($\sigma \equiv$ surface tension), so that $a \sim (\sigma/(g \rho_L))^{1/2}$.  Adopting a similar estimate for the protolunar disk, and taking $\sigma \approx 200 \, {\rm dyn}\, \cm^{-1}$, typical for molten glasses at 1 bar, $\rho_L \approx 3 \gm \cm^{-3}$, and $g \approx \Omega^2 H = 140 T_5^{1/2} x^{-3/2} \cm^2 \sec^{-1}$, implying $a \approx 1 x^{3/4} T_5^{-1/4} \, \cm$.\footnote{Drop size increases toward the midplane, since $a \propto g^{-1/2} \propto z^{-1/4}$.}  Then for reference disk parameters $\Omega \tau_{sett} \rightarrow 480 T_5^{3/8}$.  For a well-mixed disk with $f \sim 1/2$ and temperature close to liquidus, then, settling can provide as much heat as $\alpha \sim 10^{-3}$ for one settling time.

\subsection{Cooling and opacity}

The cooling time for a vapor disk is 
\begin{equation}
\Omega \, t_{cool} \simeq \frac{\Sigma c_s^2}{2\sigma T_{ph}^4} = 10^4 \,\, T_5 \left(\frac{T_{ph}}{2000\K}\right)^{-4}
\end{equation}
where $\Sigma c_s^2$ estimates the thermal energy content of the disk, and $T_{ph} \equiv$ photospheric effective temperature.   If the disk is in a steady state, the accretion rate $\dot{M}$ is known, and heating from dissipation of turbulence balances cooling then $2\sigma T_{ph}^4 = (3/(4\pi)) G \Mea \dot{M} r^{-3}$. It cannot be assumed that heating balances cooling, however, when the disk is younger than a cooling time, as is likely for the protolunar disk \citep{Charnoz2015}.  What then is the cooling time?

To evaluate $t_{cool}$ we need $T_{ph}$, but cannot assume that $T_{ph} \simeq 2000\K$ as did \cite{Charnoz2015} and \cite{Thompson1988}, motivated by the idea that this is close to the condensation temperature for solids.  We show below that energy cannot be transported out of the disk interior rapidly enough to sustain this temperature, so the disk is likely to form an opaque, cool atmosphere consisting of a mixture of solids, liquids, and cool vapor formed from volatile contaminants.  However, even if $T_{ph} = 2000\K$, $\Omega\, t_{cool} \simeq 1\times 10^{4}\, T_5^{-3}$, so independent of atmospheric structure the cooling time is long compared to the dynamical time. 

If the disk is radiative (not convective) then the usual estimate for a thin disk \citep{Hubeny1990} is
\begin{equation}
T_{ph}^4 - T_{irr}^4 \simeq \frac{3}{8} \frac{T^4}{\tau},
\end{equation}
where optical depth $\tau \simeq \Sigma \kappa$, $\kappa \equiv$ Rosseland mean opacity, and $T_{irr}$ is the effective temperature of the external radiation field.  To go further we need to know the opacity of vaporized moonrock.  

\begin{figure}[ht]
\begin{center}
\includegraphics[width=6in]{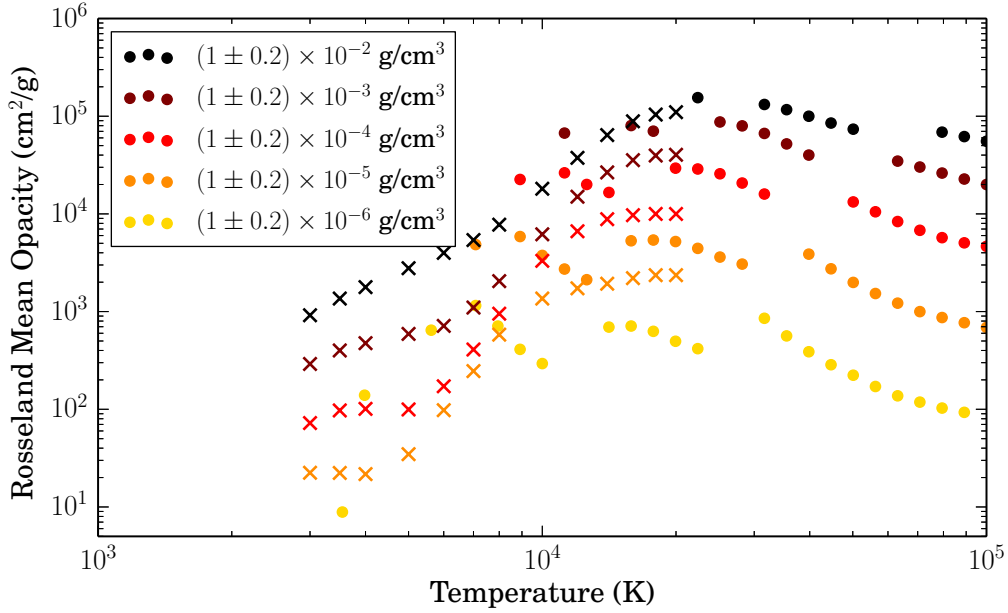} 
\caption{Rosseland mean opacity of rock vapor color coded by density. Circles are data from OP project using atomic lunar soil composition listed in Table \ref{T:composition} that do not include molecules; cross represents the calculation from \cite{Park2013} with H-chondrite composition, that do include  molecules.}
\label{fig:op}
\end{center}
\end{figure}

Figure \ref{fig:op} shows two estimates for $\kappa$.  The circles show $\kappa$ from the OP project \citep{OP} using the lunar soil composition listed in Table \ref{T:composition} \citep{Prettyman2006}; lunar soil is expected to have somewhat less iron than the bulk Moon  due to differentiation, so the opacity of mean lunar composition material is likely higher.  OP opacities are valid for $3 \times 10^3 \K < T < 10^7 \K$ and $10^{-15} \gcm < \rho < 10^{-2} \gcm$, but notice that only atomic opacities are included.   The crosses show $\kappa$ from \cite{Park2013}, which assumes H-chondrite composition and includes molecules with equilibrium abundances for $3 \times 10^3\K < T < 2 \times 10^4 \K$ and $10^{-5} \gcm  \rho < 10^{-2} \gcm$.  The difference between the two opacity estimates is less than an order of magnitude.  

In our fiducial disk model $\kappa(\rho = 10^{-3} \gcm, T_5 = 1) \simeq 590 \cm^2\gm^{-1}$, so $\tau \simeq 7 \times 10^8$.    Then $T_{ph}^4 - T_{irr}^4 \sim (30 T_5^{9/16} m^{-3/8})^4$
and
\begin{equation}
\Omega\, t_{cool} \rightarrow 2.0 \times 10^{12} T_5^{-5/4}
\end{equation}
The disk photosphere is only slightly warmer than its surroundings.  The disk is opaque, and disk cooling in inhibited by inefficient heat transport.

The disk cooling time can be reduced by convection.\footnote{In the dense protolunar disk the molecular mean free path is short, $\lambda_{mfp}/H \rightarrow 2.5 \times 10^{-15}$, thus thermal conduction is also ineffective.}  This problem has been considered by \cite{Rafikov2007} in the context of gravitational instability in protoplanetary disks.  Is the disk convective?  The condition for convective instability in a homologously contracting disk is  $\nabla_0 > \nabla_{ad}$ \citep{Lin1980,Rafikov2007}, where $\nabla_0 \equiv (1 + d\ln\kappa/d\ln p)/(4 - d\ln\kappa/d\ln T)$ and $\nabla_{ad} \equiv (\gamma - 1)/\gamma \approx 0.29$ for $\gamma = 7/5$.   A power-law fit to the \cite{Park2013} opacities near our reference model gives 
\begin{equation}
\kappa \approx 550 \left(\frac{\rho}{10^{-3}}\right)^{0.61} T_5^{1.2} \, \cm^2 \gm^{-1}.
\end{equation}
Then $\nabla_0 \approx 0.48$, and thus for the reference model the disk is convective near the midplane.  

Nevertheless, there is an upper limit to the convective heat flux.  This is set by the rate at which heat can be transported to the photosphere, where 
\begin{equation}\label{eq.photocond}
\tau = \frac{2}{3} = \int dz\, \rho \kappa \simeq \kappa_{ph} \rho_{ph} H_{ph}
\end{equation}
where quantities subscripted with $ph$ are evaluated at the photosphere, $\kappa \equiv$ opacity and $H_{ph} \simeq c_{s,ph}^2/(\Omega^2 H)$ the photospheric scale height.  The convective heat flux is approximately $\rho_{ph} c_{s,ph}^3 \mathcal{M}^3$ 
where $\mathcal{M}$ is the Mach number of turbulence at the photosphere.  We assume $\mathcal{M} < 1$.  Then 
\begin{equation}\label{eq.photobal}
\rho_{ph} c_{s,ph}^3 \mathcal{M}^3 = \sigma T_{ph}^4.
\end{equation}
Using (\ref{eq.photocond}) and (\ref{eq.photobal}),
and setting $c_s^2 = k T/\mu$, 
\begin{equation}
T_{ph} = \left(\frac{\mathcal{M}^3 k T^{1/2} \Omega}{\kappa_{ph} \mu \sigma}\right)^{2/7} \rightarrow  320\,\, \mathcal{M}^{6/7} T_5^{1/7}\kappa_{ph}^{-2/7}\K
\end{equation}
Then $\mathcal{M} < 1$ implies
\begin{equation}\label{eq.convcool}
\Omega\, t_{cool} > 1.4 \times 10^7 \,\, \kappa_{ph}^{8/7} T_5^{3/7}
\end{equation}
or of order $2000$ yr.  Shorter cooling times require supersonic convection or implausibly low photospheric opacity ($\kappa_{ph} \ll 1 \cm^2 \gm^{-1}$).  The main physical point is that the photosphere must have low density, and this limits the convective heat flux.  

A better estimate of $t_{cool}$ would model the full disk vertical structure including what could be multiple radiative and convective layers.  Cool layers close to the surface will be below solidus ($\sim 1200\K$), but we have assumed the disk can still support an atmosphere consisting of outgassed volatile vapor.  Modeling this atmosphere is an interesting and difficult problem, but beyond the scope of this paper.

\begin{table}[h] 
\centering 
\begin{tabular}{|c|c|c|c|c|c|c|c|c|c|} 
\hline
Element & O & Na & Mg & Al & Si & K & Ca & Ti & Fe \\  
\hline
Mass Fraction (\%) & 43 & 0.30 & 5.5 & 9.0 & 21 & 0.10 & 8.6 & 1.5 & 10 \\
\hline
Number Fraction (\%) & 61 & 0.29 & 5.1 & 7.5 & 17 & 0.058 & 4.8 & 0.71 & 4.0 \\
\hline
\end{tabular}
\caption{Composition of lunar soil by averaging all the columns from Apollo and Luna missions listed in Table 1 in \cite{Prettyman2006}}
\label{T:composition}
\end{table}

\subsection{Thermal evolution}

Turbulent angular momentum transport is inevitably associated with dissipation of turbulence and heating.  If we parametrize turbulent transport via Shakura-Sunyaev $\alpha$ parameter,
\begin{equation}
\Omega\, t_{heat}  = \alpha^{-1}.
\end{equation}
On the other hand, the cooling time is long.  Using the limit (\ref{eq.convcool}) heating can balance cooling only if
\begin{equation}
\alpha < (\Omega\, t_{cool})^{-1} < 7.1 \times 10^{-8} \, \kappa_{ph}^{-8/7} T_5^{-3/7}.
\end{equation}
If this condition is not satisfied, as seems likely,  then the disk will undergo runaway heating.  Unless the virial temperature $T < T_{vir} \equiv G \Mea \mu/(3 k r) = 1.1 \times 10^5 (\mu/\mu_{SiO}) x^{-1}\, \K$ is below liquidus, the disk will  vaporize completely.

\section{Disk coupling to the magnetic field}

Is a vapor disk well coupled to the magnetic field?  To answer this we evaluate the magnetic Reynolds number 
\begin{equation}
Re_M \equiv \frac{c_s H}{\eta}
\end{equation}
where $\eta = c^2/(4\pi\sigma) \equiv$ magnetic diffusivity (units $\cm^2 \sec^{-1}$), $\sigma = n_e e^2/(m_e \nu_c) \equiv$ conductivity, $n_e \equiv$ electron number density, $e \equiv$ elementary charge, $m_e \equiv$ electron mass, and $\nu_c$ is the sum of the electron-neutral and electron-ion collision frequency, all in cgs-gaussian units.  If $Re_M \gg 1$ then the field decay time is long compared to the dynamical time.  $Re_M$ is independent of the field strength.   

What are the electron-neutral and electron-ion collision frequencies?  Electron-neutral: $\nu_{c,e-n} = n_n \< \sigma v_e \> $, where $\sigma \simeq 5 (\pi a_0)^2 \simeq 10^{-15} \cm^2$ \citep{Draine2011}, and $\<v_e\> = (8 k T/(\pi m_e))^{1/2} = 4.4 \times 10^7 T_5^{1/2}\, \cm \sec^{-1}$.  Electron-ion:  $\nu_{c,e-i} = n_e (8\pi e^4 \ln\Lambda)/(m_e^2 v_e^3)$; here $\ln\Lambda \equiv$ Coulomb logarithm $\simeq 4$.  Now, $\nu_{c,e-i}/\nu_{c,e-n} = 1.7 \times 10^3 y T_5^{-2}$ where the ionization fraction $y \equiv n_e/n_{tot}$.  To evaluate this we must know $y$.

The Saha equation for the ionization fraction $X_s$ of species $s$ is
\begin{equation}
\frac{X_s^2}{1 - X_s} = \frac{2 g_{+,s} (2 \pi m_e k T)^{3/2}}{g_{n,s} \, f_s \, n_{n,s} \, h^3} e^{-\chi_s/(k T)}
\end{equation}
where $\chi_s \equiv$ ionization potential, $f_s$ is the fractional abundance by number, and $g$ are statistical weights.  For Na, $\chi_{Na} = 5.14\eV$, the ratio of statistical weights is $1/2$, and assuming $f_{Na} = 0.003$, $T_5 = 1$, and $x = x_0 = 3$; then $X_s = 0.28$ and $y > 8.5 \times 10^{-4}$.  Ionization of other atoms and molecules, especially K and Mg, increase the electron fraction by a factor of order unity.  Indeed, equilibrium models for H-chondrite vapor from \cite{Park2013} (which do not include Na) show $y \simeq 10^{-4}$ at $5000\K$; equilibrium models of \cite{Visscher2013} show $y \simeq 10^{-4}-10^{-3}$ at $T > 3000\K$ mostly from ionization of Na; the ionization fraction in \cite{Carballido2016}, based in part on \cite{Visscher2013}, also shows $y \simeq 10^{-4}-10^{-3}$, mostly from ionization of Na and K.  Hence $\nu_{c,e-i}/\nu_{c,e-n} \gtrsim 1$ at $T \sim 5000\K$.

Combining estimates, 
\begin{equation}
Re_M = 4.9 \times 10^5 T_5^{5/2} x^{3/2} \qquad 
y > 5.7 \times 10^{-4} \,  T_5^2
\end{equation}
\begin{equation}
Re_M = 8.5 \times 10^8 T_5^{1/2} x^{3/2} y \qquad 
y < 5.7 \times 10^{-4} \,  T_5^2
\end{equation}
where in the former case electron-ion collisions dominate, and $Re_M$ is independent of the ionization fraction.

\begin{figure}[ht]
\includegraphics[width=6in]{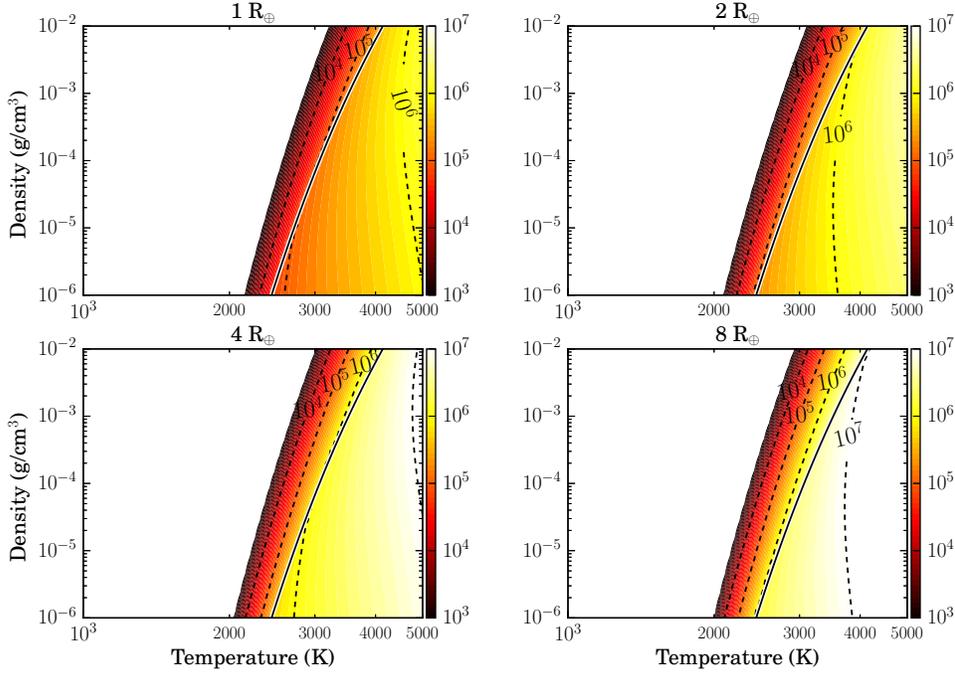} 
\caption{Magnetic Reynolds number ($Re_M$) calculated at a range of radii, density, and temperatures below the virial temperature. $Re_M \gg 1$ for $T \gtrsim 3500 \K$, which indicates good coupling between a vapor disk and the magnetic field. The solid line is an estimate for the vaporization temperature; see text for details.  For temperatures below the vaporization temperature we set the conductivity to be the conductivity of the vapor phase.}
\label{fig:ReM}
\end{figure}

Figure \ref{fig:ReM} shows an estimate of $Re_M$ that uses $\nu_c = \nu_{c, e-n} + \nu_{c, e-i}$ and $n_e$ from the Saha equation with lunar soil composition (Table \ref{T:composition}).  The estimate assumes all elements are in atomic form.  A full equilibrium calculation would be valuable but is beyond the scope of this paper. The four panels in Figure \ref{fig:ReM} show $Re_M$ at different radii within a range of density and temperature (which must however be below  $T_{vir}$). Evidently $Re_M$ is $\gg 1$ where $T \gtrsim 4000 \K$.  

We have assumed a vapor disk, but at sufficiently low temperature the disk will consist of a two-phase liquid-vapor mixture \citep{Thompson1988}.  How far into the mixed regime is the disk well-coupled?  

Consider a two-phase homogeneous medium at the vaporization temperature, estimated to lie at $P = 9.4 \times 10^{13} \exp(-11.4/T_5) {\rm dyn} \cm^{-2}$ \citep[][but see Visscher and Fegley]{Melosh2007}.  In the reference disk this corresponds to $T \simeq 4000\K$, $\rho \simeq 7 \times 10^{-3}\gm \cm^{-3}$, and liquid/vapor density contrast $\simeq 430$, assuming $\rho_L \simeq 3 \gm \cm^{-3}$.  Adopting our earlier estimate for liquid droplet size $a \sim 1 \cm$, the number density of droplets is small and they are well separated unless $f \sim 1$.  If we assume the electrical conductivity of the droplets is lower than the vapor, then electrical currents will flow through the vapor, which is connected and occupies most of the volume.  The conductivity will then be determined by the electron abundance and electron collision frequency in the vapor phase. 

Figure \ref{fig:ReM} incorporates an estimate of $Re_M(\rho,T)$ in this regime, assuming that the composition of the liquid and vapor phases are identical.  This is a conservative assumption, since Na is the main electron donor and may be concentrated in the vapor phase.  Evidently the conductivity in a two-phase disk is determined mainly by the ionization state of Na vapor and therefore by the temperature.  Close to liquidus $Re_M$ drops rapidly and the disk begins to decouple.  The locus $Re_M(\rho,T) = 10^4$ is well fit by 
\begin{equation}
T_5^{-1|} = 1.04 - 0.084 \ln\rho + 0.091 \ln x\qquad \Leftrightarrow \qquad Re_M = 10^4
\end{equation}
where $\rho$ is in cgs, and recall that $x$ is radius in units of $\Rea$.  Here the radius enters only through the requirement that the disk is in vertical hydrostatic equilibrium, and the reference model has not been used.

We have considered the effect of finite conductivity (Ohmic diffusion), but in  protostellar disks the Hall effect and ambipolar diffusion are known to be as or more important nonideal effects  \citep[see][and references therein]{Turner2014}. Are the Hall effect and ambipolar diffusion important in the protolunar disk?  

Beginning with the discussion of \cite{Balbus2001}, the
ratio of the Hall to Ohmic term in the induction equation $\sim \omega_{c,e}/\nu_{c,e-i} \simeq 2.4 \times 10^{-4} T_5^{9/4} m^{-1/2} y^{-1} \beta^{-1/2}$.  Here $\omega_{c,e} \equiv e B/(m_e c)$, $\beta$ is the ratio of gas to magnetic pressure and we have assumed that the $\nu_{coll,e} = \nu_{c,e-i}$.  For our estimated $y$ and $\beta \gtrsim 1$ the Hall effect is at most comparable to Ohmic diffusion.  

Similarly, the ratio of the ambipolar to Ohmic term $\sim (\omega_{c,i}/\omega_{c,e}) (\nu_{e,n}/(\gamma_d \rho)) \simeq 8.0 \times 10^{-4} T_5^{1/2}$, where $\gamma_d \simeq 1.7 \times 10^{13} \cm^3 \sec^{-1}\gm^{-1}$ is the drag coefficient, assuming Mg ions and a neutral gas of SiO \citep{Draine1983}.  For $T_5 \sim 1$, then, ambipolar diffusion is much less important than Ohmic diffusion.  The importance of ambipolar diffusion and the Hall effect would need to be reevaluated for conditions very different from our reference state, including in conditions close to the disk surface.

\section{Implications of magnetic coupling}

Evidently $Re_M \gg 1$ in the vapor disk from close to the Earth's surface to some outer radius where the disk is too cool to couple. What are the consequences?

The magnetic field is dynamically significant if it is close to equipartition: $B \sim \sqrt{8\pi p} = 1.7 \times 10^4 m^{1/2} T_5^{1/4}$G.  The magnetic field strength of the early Earth is not known \citep[e.g.][]{Tarduno2015}.  Even field strengths as high as tens of kilogauss in the  vapor disk would leave no trace, however, if the disk later cooled, decoupling the field and allowing it to escape before formation of solids.  Although the pre-impact magnetic field is unlikely to be close to equipartition, turbulence can amplify an initially weak field until it is close to equipartition with the turbulent kinetic energy \citep[e.g.][]{Meneguzzi1981}.  What are the potential sources of turbulence?

In ideal MHD differentially rotating disks are subject to the magnetorotational instability \citep[MRI][]{Balbus1991}, and the MRI saturates in a turbulent state \citep[e.g.][]{Hawley1995}.  At finite resistivity, however, the growth rate depends on $Re_M$ and the field strength, and falls into one of three regimes:  

(1) If the Alfv\'en speed $V_A \equiv B/(4\pi\rho)^{1/2} > Re_M^{-1/2} c_s$, or $B \gtrsim 10 \, \G$, the MRI has maximum growth rate $\simeq (3/4)\Omega$ (here $\Omega = \sqrt{ G M/r^3} \equiv$ orbital frequency).   Saturation of the MRI in this regime is not yet fully understood, and may depend on the magnetic Prandtl number $Pr_M \equiv \nu/\eta \sim 6.4 \times 10^{-8} T_5^{5/2}$ ($\nu \equiv$ kinematic, not turbulent, viscosity), although recent numerical evidence suggests the dependence vanishes at low $Pr_M$ for sufficiently high Reynolds number $Re$; we estimate $Re \sim 4 \times 10^{13}$.  Most high resolution numerical experiments show exponential growth of the field strength saturating at $\beta \sim  20$, as do high resolution global disk simulations \citep[e.g.][]{Shiokawa2012,Hawley2013}.

(2) If $Re_M^{-1/2} c_s > V_A > Re_M^{-1} c_s$ (the latter limit corresponds to $\sim 6$mG in the reference disk) the MRI is still present but the maximum growth rate is reduced to $\sim V_A^2/\eta$.  Simulations at modest $Re_M$ suggest that the growth of the field is weakened or halted in this regime, and that the outcome depends on the magnetic Prandtl number \citep{Turner2014}.  

(3) If $V_A < Re_M^{-1} c_s$ the field is still trapped in the disk (so long as $Re_M > 1$) but the MRI is suppressed.  The field is amplified by turbulence as long as the Ohmic diffusion time across a turbulent eddy is longer than the eddy turnover time.  For a convective disk where the turbulent eddies have a scale of order $H$ the field would be amplified if $Re_M$ is larger than the inverse Mach number of the convection.  

Our best-bet scenario is that once the disk is hot enough to couple to the magnetic field, an initially weak field would be amplified by turbulence in the disk (provided, for example, by convection) until it is strong enough that MRI driven turbulence can lift off.  Then the field would be amplified to slightly sub-equipartition levels and full-blown MHD turbulence would drive disk evolution.

Once MRI is active, the heating timescale $\Omega\, t_{heat} \sim \alpha^{-1}$.  Numerical studies of MHD turbulence in disks give $\alpha \sim 0.03$ \citep[e.g.][]{Turner2014,  Ryan2016}, so $t_{heat} \simeq 46$hrs in our fiducial model.  The disk evolution timescale $t_{spread} \Omega = \alpha^{-1}(R/H)^2 = 16 \alpha^{-1} T_5^{-1}$, so $t_{spread} \simeq 600$hr. The shortest timescale for the MRI-active disk is the dynamical time, followed by the heating time, followed by the disk spreading time.  The cooling time is likely so much longer than all these timescale that cooling can be completely neglected.

\section{Scenario for Disk Evolution}

Let us now suppose that most of the mass and angular momentum in the disk is at a few Earth radii as in our reference model.  How might the post-impact disk evolve?  

If the initial disk is cool enough to be decoupled from the magnetic field, internal turbulence and shock waves generated in the boundary layer heat the disk.  As long as the heating timescale is longer than the cooling timescale, $Re_M$ increases and eventually the disk is well coupled to the magnetic field. 

Next, assuming a sufficiently strong seed field, the MRI takes off and drives a turbulent state with $\alpha \sim 10^{-2}-10^{-1}$.  Since cooling is ineffective the disk undergoes runaway heating until it reaches the virial temperature $T_{vir}$, equivalent to $H/r \sim 1$ and $t_{spread} = (1/(\alpha\Omega))(r/H)^2 \simeq t_{heat} = 1/(\alpha\Omega)$.  The disk, which in the reference model begins as effectively a ring of material at $r \simeq x_0 \Rea$, spreads, cooling adiabatically at its outer boundary and accreting onto Earth at its inner boundary.  The geometrically thick disk resembles the radiatively inefficient accretion flow (RIAF) model used in black hole accretion studies \citep{Yuan2014}.  

If there is any cool material left at the disk midplane, it will likely accrete onto Earth once the disk thickens.  Thick disks orbit at sub-Keplerian speeds, so solid bodies embedded in the vapor disk and orbiting at the Keplerian velocity face a stiff headwind.  A thin liquid disk will exchange angular momentum with the overlying vapor disk through a turbulent boundary layer.  Provided that the bulk of the disk is vapor and that the solid bodies are not too large they can also be expected to lose angular momentum to the disk and accrete.  This point was also made by \cite{Carballido2016}.

Nearly half the dissipation in disk accretion occurs at the boundary layer between disk and central object.  It is now believed that the transition through the boundary layer is mediated by torques from compressive waves and shocks rather than magnetic fields \citep{Belyaev2012b, Belyaev2013a, Belyaev2013b} \footnote{The shear in the boundary layer may nevertheless amplify the magnetic field.}.  This deposits high entropy material at the boundary layer, possibly driving convection.  

High entropy vapor generated directly by accretion through the boundary layer or by dissipation of shocks in the disk atmosphere may become unbound in the sense that the Bernoulli parameter  $Be \equiv h + v^2/2 + \phi > 0$ ($h \equiv$ enthalpy, $\phi \equiv$ gravitational potential).  The boundary layer might then source a powerful, magnetized outflow carrying away mass and angular momentum.  

T Tauri stars, for example, have long been thought to rid themselves of excess accreted angular momentum through wind-mediated magnetic braking \citep{Hartmann1989}. In that case, where the disk contains an enormous reservoir of mass and angular momentum, the star reaches spin equilibrium with accreted angular momentum balanced by wind angular momentum losses: $\dot{M}_{acc} l_{acc} \simeq \dot{M}_{w} l_{w}$, where $l_{acc}$ is the specific angular momentum of the accreted matter and $l_w$ is the specific angular momentum of the wind.  If the wind originates near the stellar surface $l_w/l_{acc} \simeq (r_A/r_*)^2$, where $r_A$ is the Alfv\'en radius, which depends on the dipolar field strength at the stellar surface \citep[see][for a discussion]{Matt2008}.  Evidently if $r_A/r_* \gtrsim $ a few, then the magnetized wind can sharply change the angular momentum but not the mass budget.  

In the Earth-protolunar disk system the disk contains a relatively small fraction of the total mass and angular momentum, so it is less clear that spin equilibrium can be achieved.  Nevertheless, if the boundary layer drives a wind, the early Earth is rapidly rotating, the boundary layer and disk have a strong well-organized field, and the wind is sufficiently ionized to couple to the magnetic field, then angular momentum could be efficiently removed by the wind and the angular momentum constraint on the giant impactor could be lifted completely.

How rapidly does the disk spread, and how rapidly does matter accrete?  We can assess this using a simple model inspired by the similarity solution of \cite{Ogilvie1999}.  First, notice that most of the reference disk's mass $M_D$ is at its outer edge: $d M_D/d\ln r = d (\Sigma \pi r^2)/d\ln r = M_D (2 + d\ln\Sigma/d\ln r) > 0$.  Mass is concentrated at the outer edge in any disk in which $d\ln\Sigma/d\ln r > -2$, and this is the case in, for example, the ADAF model, which has $\Sigma \sim r^{-1/2}$.  The disk is therefore effectively a ring of radius $r_D$ with total angular momentum $J_D \simeq M_D \sqrt{G \Mea r_D}$.  

The evolution of the disk/ring can be modeled by
\begin{equation}
\frac{d M_D}{dt} = - \alpha h^2\Omega M_D + \dot{M}_{ext}
\end{equation}
where $h \equiv (H/r)$ and $\dot{M}_{ext} \equiv$ models any outflow or inflow other than accretion onto Earth.  The accretion rate estimate comes from the usual $\alpha$ disk estimate $\dot{M} \sim \Sigma \nu = \Sigma \alpha c_s^2/\Omega$, $c_s^2/(\Omega^2 r^2) \sim h^2 \sim 1$ (thick disk), and $\Sigma \sim M_D/r^2$.  Angular momentum conservation gives 
\begin{equation}
\frac{d J_D}{dt} = \frac{d}{dt}( M_D \sqrt{G \Mea r_D} ) + \tau_{ext}
\end{equation}
where $\tau_{ext}$ models external torques.  If $\tau_{ext} = 0$ and $\dot{M}_{ext} = 0$, then the model admits the solution
\begin{equation}
r_D = r_0 (1 + t/t_0)^{2/3}
\end{equation}
and
\begin{equation}
M_D = M_0 (1 + t/t_0)^{-1/3}
\end{equation}
subject to the initial conditions $r_D(t = 0) = r_0$ and $M_D(t = 0) = M_0$.  Here $t_0 \equiv (3 \alpha h^2 \Omega (r_0))^{-1}$.

The outer edge of the disk spreads and cools, and the disk loses mass at the inner edge.\footnote{The disk entropy evolution follows from  $\rho \sim M_D/r_D^3$, $p \sim \rho G \Mea/r_D$, so $p/\rho^\gamma \sim t^{7\gamma/3 - 3}$, where $\gamma$ is the (assumed constant) adiabatic index.}  If the disk midplane temperature is $T = h^2 T_{vir} = 7.5 \times 10^4 x^{-1} h^2 (\mu/30)$K, it will reach a critical temperature $T_{crit} \sim 4000K$ for decoupling when either (1) the cooling time is comparable to the spreading time, or (2) when the disk reaches a radius where $Re_M(\rho, T_{vir})$ is small enough for decoupling, i.e. at $r_D \simeq 40 h^2 \Rea$.  Which process initiates decoupling depends sensitively on $h$, disk evolution, and disk thermal physics.  

Suppose that $h = 1/2$. Then decoupling occurs at $r_D \simeq 10 \Rea$ when the disk surface density is $\sim 5 \times 10^5 \gm \cm^{-2}$ and the radiative cooling time, at least, is still $\gg (\alpha h^2 \Omega)^{-1}$.  The disk reaches this radius at $t = 160 (\alpha/0.05)^{-1} \,$hrs, when about one third of the original disk has accreted.  In our scenario, the resulting decoupled vapor cools and provides the raw material for formation of the moon. 

\section{Discussion and Conclusion}

In this paper we have investigated the early evolution of a remnant disk formed by a giant impact with Earth. 

We estimated that a vapor disk has large Rosseland mean optical depth.  Cooling is ineffective, even if the disk is convective.  Any form of turbulent angular momentum transport, characterized by Shakura-Sunyaev parameter $\alpha$, will heat it on a timescale $(\alpha\Omega)^{-1}$, which is short compared to the thin disk evolution timescale $(\alpha\Omega)^{-1} (r/H)^2$.  

We showed, following \cite{Carballido2016}, that if the disk contains a vapor component with $T \gtrsim 4000\K$ then that component is well coupled to the magnetic field.  The precise lower limit for coupling depends on composition, particularly the abundance of K and Na.  Once the critical temperature is exceeded--and this may happen during initial collision--then there is the possibility of MHD turbulence driven by the magnetorotational instability.  

The evolution of a magnetically coupled disk depends on the initial field strength and geometry.  If MHD turbulence is present, the numerical evidence suggests that it will heat the disk still further and transport angular momentum efficiently.  

Assuming that angular momentum {\em is} transported efficiently, we have put forward a scenario in which the disk first heats to the virial temperature and then spreads on a timescale of $\sim 600$ hrs.  The outcome is a ring of material at the outer edge of the disk that spreads and cools until it decouples from the magnetic field.  Using a simple model, we estimate that the decoupled remnant disk has radius $\sim 10 \Rea$.  We estimated that the disk mass $\sim t^{-1/3}$, and that by the time the disk decouples $\sim 1/2$ of the original disk mass is left.  

A large fraction of the protolunar disk's power is dissipated in the boundary layer where the disk meets Earth's surface.  The boundary layer will produce high entropy material, and we have speculated that this material might mix back into the disk, or become unbound and leave in the form of a powerful, possibly magnetized wind originating from the Earth or from the disk itself \citep{Blandford1982}.  Any wind from the disk is likely enhanced in volatiles: inefficient heat transport from the disk interior implies that the disk surface temperature cannot be sustained above the grain condensation temperature, so the disk will outgas as condensation and settling are driven by radiative cooling at the surface.

Two uncertainties hang over our scenario.  First, what is the distribution of temperatures in the post-impact disk?  Most numerical simulations of the collision generate some hot material in the disk, with the final temperature distribution dependent on initial conditions.  Nevertheless, even if the inital disk is cool, any angular momentum diffusion in the post-collision disk will heat the disk, and the inefficiency of heat transport guarantees that the disk will heat before it spreads, reaching $Re_M \gg 1$.   Second, what is the initial field strength?  If the field is weak enough then resistive diffusion damps the magnetorotational instability, and (if $Re_M > 1$) differential rotation will provide only a modest, linear-in-time field amplification.  

The post-impact temperature distribution depends on details of the impact dynamics.  Simulations of merging magnetized neutron stars \citep[e.g.][]{Kiuchi2015}--also a merger of degenerate objects--exhibit fields that are amplified by at least a factor of $10^3$ in turbulence driven by shear discontinuities formed in the collision.  In \cite{Kiuchi2015} the amplification increases with resolution, with no sign of convergence.  It is reasonable to think that the field will saturate when  magnetic energy is comparable to turbulent kinetic energy, and requires only a few shear times, a time comparable to the duration of the collision.  In sum, it is plausible that magnetic coupling alters the dynamics of the collision itself, the subsequent circularization of the disk, and the initial thermal state of the disk. 

The magnetic field strength and geometry of the pre-impact Earth and impactor will likely never be known.  Still, one can ask how weak a field is required to initiate runaway heating of the disk.  The boundary layer may be particularly constraining because $\sim 0.3 \eV$ per nucleon is dissipated in the layer, suggesting that the boundary layer will immediately generate  hot, well-coupled vapor even if coupling is initially poor elsewhere in the disk.  Turbulence associated with the boundary layer might then provide a large amplification factor for the initial field and mix it outward into the disk.

Our work follows the recent interesting paper by \cite{Carballido2016} (hereafter CDT), who demonstrate that the protolunar disk is likely to be magnetically coupled (\cite{Charnoz2015} also suggested that the protolunar disk might be well coupled, but they do not provide a detailed evaluation of the ionization fraction or instability conditions).  CDT also consider mixing in the protolunar disk.  We have performed a less careful evaluation of the ionization fraction, but CDT's work suggests that ionization is, in any event, dominated by Na and K.  While CDT use an unstratified shearing box model to estimate a lower limit on the angular momentum transport efficiency due to MHD turbulence of $\alpha \sim 7 \times 10^{-6}$, these zero-net-flux, unstratified, shearing boxes are known to be nonconvergent \citep{Fromang2007}.  Simulations of stratified shearing box models, global models, and models with explicit dissipation tend to produce $\alpha \sim $few$ \times 10^{-2}$.  The weight of numerical evidence therefore suggests much higher $\alpha$ and more rapid evolution of a magnetized protolunar disk.  

Interestingly, stratified shearing box models \citep[e.g.][]{Stone1996, Davis2010, Ryan2016} show that $\alpha$ depends on distance from the midplane, with $\alpha \sim 1$ at $z \sim 2 H$.  If this obtains for a near-virial protolunar disk, and the ratio of turbulent angular momentum diffusion to turbulent mixing is of order unity \citep{Carballido2005}, then mixing would occur on a small multiple of the {\em dynamical} timescale.  The same efficient mixing might also transport magnetic fields outward from the boundary layer into the bulk of the disk.

\cite{Charnoz2015} (hereafter CM) recently considered several scenarios for the long-term evolution of the protolunar disk with the aid of a numerical model.  CM's models typically have a hot disk near the inner edge, close to the boundary layer, with $T \sim 5000$K, consistent with magnetic coupling.  Following \cite{Thompson1988} and others, CM assume that the disk cools from the surface with a photospheric effective temperature $\sim 2000$K.  CM do not solve self-consistently for the temperature of the disk photosphere, although this is exceedingly difficult because a cool surface would consist of a mixture of vapor, liquids, and solids.  As in CM, viscous heating and cooling do not balance in our scenario.

Our scenario is unorthodox in that it assumes a hot initial disk, with the cold disk forming later at of order ten Earth radii.  In the canonical picture the moon forms just outside the Roche radius.  The early evolution of the Moon's orbit is very poorly constrained, however.  Certainly the tidal coupling of Earth and Moon is too poorly known for this to constrain the initial semimajor axis of the moon \citep{Bills1999}.

\acknowledgments
CFG's work was supported in part by a Romano Professorial Scholar appointment, a Simons Fellowship in Theoretical Physics, and a Visiting Fellowship at All Souls College, Oxford.  We thanks S. Desch, C. Thompson, M. Chandra, B. Ryan, J. Papaloizou, and C. Terquem for comments, and the referee, S. Charnoz, for a thoughtful report that greatly improved the manuscript.


\end{document}